\documentstyle[12pt]{article}
\date{}
\begin{document}

\title{The Breathing Modes of the $B=2$ Skyrmion and the Spin-Orbit
Interaction}

\author{A. P. Kobushkin$^1$ and D. O. Riska$^2$}
\maketitle

\centerline{\it $^1$Bogolyubov Institute for Theoretical Physics}
\centerline{\it National Academy of Sciences of Ukraine,
                 252143 Kiev, Ukraine}

\centerline{\it $^2$Department of Physics, University of Helsinki,
00014 Finland}

\setcounter{page} {0}
\vspace{1cm}

\centerline{\bf Abstract}
\vspace{0.5cm}

The coupling of the breathing and rotational modes of the
skyrmion-skyrmion system leads to a
nucleon-nucleon spin-orbit interaction
of short range, as well as to spin-orbit
potentials for the transitions
$NN \to N(1440)N$, $NN \to NN(1440)$ and
$NN \to N(1440)N(1440)$.
The longest range behaviour of
these spin-orbit potentials is calculated in closed
form.\\

Preprint HU-TFT-96-57

\newpage

The Skyrme model, and its generalizations, provide theoretically
compact models for the baryons, which are able to describe most of
their observed low energy properties to an accuracy of 30\% or
better \cite{Skyrme61,ANW}. By quantization of the appropriate collective
coordinates, the Skyrme model provides a natural and conceptually
simple method for describing the rotational and vibrational
degrees of freedom, that are revealed by the baryon spectrum. As
an example the global adiabatic rotation of the skyrmion generates
the spin- and isospin operators for the baryons \cite{ANW}.
Furthermore employment of the so called product ansatz for
the $B=2$ skyrmion \cite{Skyrme62} makes it possible to quantize the motions
of the individual skyrmions so as to generate a well defined
nucleon-nucleon interaction operator for two spatially extended
nucleons that is asymptotically exact \cite{Skyrme62,Jackson,NymanRiska}.\\

To the important collective modes of the skyrmion also belong the
brea\-th\-ing modes, which involve spherically symmetrical size
vibrations. The breathing mode is responsible for the lowest
even parity resonances of the nucleon and the $\Delta_{33}$:
the $N(1440)$ and
$\Delta(1600)$, respectively \cite{Hajduk,BDT}.
If the breathing
mode is included as a quantum correction in the Hamiltonian it
strongly affects the soliton profile \cite{Kostyuk}. The net effect
of this additional breathing mode induced spin-orbit
interaction in the $NN$ system is small at large distances
but at short range it appears -- at least in the 
approximation below
 -- to be large enough to change the
sign of the isospin dependent
spin-orbit interaction that is induced by
the rotational modes \cite{Nyman,RiskaNyman,RiskaDannbom}.\\

We here consider the breathing modes for the two-skyrmion system
treated in the product approximation, and show
that their interplay with the
rotational degrees of 
freedom leads to an effective spin-orbit interaction
between two nucleons, concomitantly with a spin-orbit coupling between
the nucleon nucleon ($NN$) and the $NN(1440)$ and
$N(1440)N(1440)$ channels. \\

The product approximation for the $B=2$ skyrmion, which is justified
when the separation bet\-ween
the centers of the two skyrmions, $\vec R_1(t)$ and $\vec R_2(t)$ exceeds
the skyrmion size, is \cite{Skyrme62}
\begin{equation}
U\left(\vec r,t;\vec R_1(t),\vec R_2(t)\right)=
U_1\left(\vec r-\vec R_1(t);t\right)
U_2\left(\vec r-\vec R_2(t);t\right)\ .\label{PA}
\end{equation}
Here $(\vec r,t)$ is the spacetime coordinate of the soliton field and
$U_i\left(\vec r-\vec R_i(t);t\right)$ is the adiabatically rotated hedgehog
ansatz for the $i$-th skyrmion centered at $\vec R_i(t)$:
\begin{equation}
U_i\left(\vec r_i;t\right)=A_i(t)U_i^0(\vec r_i)A_i^\dagger(t),
\quad
U_i^0\left(\vec r_i\right)=exp\{i\vec \tau \cdot \hat r_iF(r_i)\}\ .
\label{CF}
\end{equation}

Here we use the abbreviation
$\vec r_i\equiv \vec r-\vec R_i(t)$, and denote
the time dependent $SU(2)$ rotation matrix for the
$i$-th skyrmion $A_i(t)$ and the chiral angle $F(r)$ \cite{ANW}.\\

For a single $(B=1)$ skyrmion the breathing mode coordinate
$\lambda(t)$ is introduced by the replacement of the static hedgehog
field $U^0(\vec r)$ by the time dependent field $U^0(\vec r,t)\equiv
U^0(\lambda(t)\vec r$).  By this scale transformation the mass of the
classical soliton
becomes a function of the breathing coordinate with its
absolute minimum at $\lambda=1$. At the quantum level the
strong coupling of the breathing and the rotational modes \cite{BDT}
shifts the minimum of the effective breathing potential $V(\lambda)$ for a
free nucleon
to $\lambda=\lambda_0=0.868$. In the vicinity of this minimum
potential $V(\lambda)$ can be expanded as
\begin{equation} V(\lambda_0+\xi)\simeq V(\lambda_0)+\xi^2
V\ ,\ \ \ \ V(\lambda_0)=\frac{F_{\pi}}{e_S}\times 39.14\ ,\ \ \ \
V=\frac{F_{\pi}}{e_S}\times 35.55.
\label{BDT-pot}
\end{equation}
For the parameter values
$F_{\pi}=129\ MeV$ and $e_S=5.45$ used in ref.\cite{ANW} one obtains
the expectation
values $<0|\xi^2|0>=0.145$ and
$<0|\xi^2|1>=0.380$, where $|0>$ and $|1>$ are the
ground and the
first excited breathing states of the $B=1$ skyrmion
respectively. \\

For the $B=2$ skyrmion considered in the
framework of the product
ansatz (\ref{PA}) the breathing modes are introduced by the substitution
$U=U_1(\lambda_1 \vec r_1,t) U_2(\lambda_2 \vec r_2, t)$
\cite{Kalbermann}. The masses
of the free solitons then become nonequal $M(\lambda_1) \neq M(\lambda_2)$
with absolute minimum at $\lambda_1=\lambda_2=\lambda_0$. In this case
the interaction between the two skyrmions with $B=1$ induces a coupling of the
$NN$-system to $NN(1440)$ and
$N(1440)N(1440)$ states, which gives rise to a repulsive
central interaction between two nucleons \cite{Kalbermann}.\\

Because the the product ansatz provides a reliable
approximation 
only for the longest range
behaviour of any component of the skyrmion-skyrmion interaction
we shall here derive only the longest range component of the spin-orbit
interaction, which is generated by the coupling of the vibrational and
rotational modes. For this it is sufficient to consider only the time
derivatives in the quadratic term in the Lagrangian density of the model.
The relevant part of the nucleon-nucleon interaction then
takes the form \cite{RiskaNyman}
\begin{equation}
V=-{F_\pi^2\over 8}\int d^3r\ Tr\left(U_1^\dagger(\vec z_1,t)\dot{U}_1
(\vec z_1,t)\dot U_2(\vec z_2,t){U}_2^\dagger(\vec z_2,t)
\right)\ , \label{Potential}
\end{equation}
where $\vec z_i=\lambda_i \vec r_i$ and the dots represent total
time derivatives.\\

Using the projection formulae \cite{NymanRiska,Nyman}
\begin{eqnarray}
&&<N_i^{\
\prime}|A_i\tau _n j_m^{(i)}A_i^{\dagger}|N_i> =-\delta _{nm} <N_i^{\
\prime}|\frac{1}{6}\vec{\tau}\cdot \vec{\tau}^{(i)}+\frac{1}{2} |N_i>
\ , \\
&&<N_i^{\ \prime}|A_i \tau _n A_i^{\dagger}|N_i>
=-\frac{1}{3}
<N_i^{\ \prime}|\sigma_n^{(i)} \vec{\tau}\cdot \vec{\tau}^{(i)}|N_i>
\label{projectile}
\end{eqnarray}
one readily finds the explicit expressions:
\begin{eqnarray}
U_1^{\dagger}\dot{U}_1&&\!\!\!\!\!\!\!\!\!
=+\frac{i}{3}\vec{\tau}\cdot \vec{\tau}^{(1)}\left[\frac{1}{I(\lambda_1)}
+\frac{(\vec{\sigma}^{(1)}\cdot
\vec R \times \dot{\vec R})(\vec r_1\cdot \vec R)}
{2 R^2 r_{1}^2}\right]\ sin^2F(z_1)+ ... \ ,
\label{U_1-dot} \\
\dot{U}_2U_2^{\dagger}&&\!\!\!\!\!\!\!\!\!
=-\frac{i}{3}\vec{\tau}\cdot \vec{\tau}^{(2)}\left[\frac{1}{I(\lambda_2)}
+\frac{(\vec{\sigma}^{(2)}\cdot
\vec R \times \dot{\vec R})(\vec r_2\cdot \vec R)}
{2 R^2  r_{2}^2}\right]\ sin^2F(z_2)+ ...\ ,
\label{U_2-dot}
\end{eqnarray}
where $\vec{\sigma}^{(i)}$ and $\vec{\tau}^{(i)}$ are the spin
and isospin Pauli matrices for the $i$-th nucleon and $I(\lambda_i)$
is its moment of inertia, the latter being a function of the breathing
coordinate $\lambda_i$ \cite{BDT}. The dots
indicate terms, which do not contribute to the spin-orbit force.\\

Substitution of the expressions (\ref{U_1-dot}) and (\ref{U_2-dot}) in
(\ref{Potential}) yields the spin-orbit interaction:
\begin{equation}
V_{\! LS}=(\vec{\tau}^{(1)}\cdot \vec{\tau}^{(2)})
\left\{
(\vec{\sigma}^{(1)}+\vec{\sigma}^{(2)}) \vec{l}\ V_{+}
+(\vec{\sigma}^{(1)}-\vec{\sigma}^{(2)}) \vec{l}\ V_{-}
\right\}\ ,
\label{V_LS}
\end{equation}
where the functions $V_\pm$ are defined as
\begin{equation}
V_{\pm}=-\frac{F_{\pi}^2}{72MR^2}\int d^3r\left[
\frac{(\vec r_1\cdot \vec R)}{r_1^2 I(\lambda_2)} \pm
\frac{(\vec r_2\cdot \vec R)}{r_2^2 I(\lambda_1)}\right]
sin^2 F(z_1) sin^2 F(z_2).
\label{I}
\end{equation}
Note the appearance of an antisymmetric spin-orbit interaction for the
case when $\lambda_1 \neq \lambda_2$.

To estimate the long range behaviour of the potential we shall use the
approximation where
the field of one skyrmion is constant in the vicinity of the other one
\cite{NymanRiska,Nyman}. This leads to the expressions
\begin{eqnarray}
V_{+} \simeq \frac{1}{12MR^2}\left[
v(\lambda_1R)+v(\lambda_2R)
+\left(\frac{\lambda_2}{\lambda_1}\right)^3 u(\lambda_2R)
+\left(\frac{\lambda_1}{\lambda_2}\right)^3 u(\lambda_1R)
\right],
\label{V_+}                   \\
V_{-} \simeq \frac{1}{12MR^2}\left[
\left(\frac{\lambda_2}{\lambda_1}\right)^3 u(\lambda_2R)
-\left(\frac{\lambda_1}{\lambda_2}\right)^3 u(\lambda_1R)
\right]
\label{V_-} \ ,
\end{eqnarray}
where we have used the abbreviations
\begin{equation}
v(R)=-sin^2F(R)\ ,  \ \
u(R)=\frac{1}{3}RF^{\prime}(R)\;sin2F(R) \ .
\label{v-potentials}
\end{equation}
In (\ref{V_+}) we have used the quadratic approximation,
\begin{equation}
I(\lambda_i)=\frac{F^2_{\pi}}{6\lambda_i^3}\int d^3r\ sin^2F(r)
\label{I} \ ,
\end{equation}
for the moment of inertia  \cite{NymanRiska}.\\

Expansion of the potentials $V_{+}$ and $V_{-}$ in the 
vicinity of the minima at
$\lambda_1 \simeq \lambda_0$ and $\lambda_2\simeq \lambda_0$ gives
\begin{eqnarray}
&&V_+ \simeq V_0 + (\xi_1+\xi_2)V_1^{+} + (\xi_1^2+\xi_2^2)V_2^{+}
+ \xi_1\xi_2 V_3^+ \ ,
\label{expanV+} \\
&&V_- \simeq (\xi_1-\xi_2)V_1^{-} + (\xi_1^2-\xi_2^2)V_2^{-} \ ,
\label{expanV-}
\end{eqnarray}
respectively. Here $\xi_i=\lambda_i-\lambda_0$ and the
potential functions $V_j^\pm$ are defined as
\begin{eqnarray}
&&V_0=\frac{1}{6MR^2}\left(-sin^2F +
\frac{1}{3} \tilde{R} F^{\prime}s\right)
\label{V_0}   \ ,                   \\
&&V_1^+=\frac{1}{36MR}\left[(-2F^{\prime}+\tilde{R} F^{\prime \prime})s
+ 2\tilde{R} {F^{\prime}}^2c\right] \ ,
\label{V_1^+}\\
&&V_2^+=\frac{1}{36MR}\left[(12F^{\prime}+4\tilde{R} F^{\prime \prime}
+\frac{1}{2}\tilde{R}^2 F^{\prime \prime \prime}-
2\tilde{R}^2 {F^{\prime}}^3 )s \right.
\nonumber \\
&& \ \ \ \ \ \ \ \ \ \ \
\left. + (4\tilde{R} {F^{\prime}}^2
-\frac{3}{2}\tilde{R} F^{\prime \prime}
+3\tilde{R}^2 F^{\prime}F^{\prime \prime})c\right] \ ,
\label{V_2^+} \\
&&V_3=-\frac{1}{12MR}\left[(8F^{\prime}
+2\tilde{R} F^{\prime \prime})s
 + (4\tilde{R} {F^{\prime}}^2c \right] \ ,
\label{V_3} \\
&&V_1^{-}=-\frac{1}{36MR}\left[(4F^{\prime}
+\tilde{R} F^{\prime \prime})s+2\tilde{R} {F^{\prime}}^2c
\right] \ ,
\label{V_1^-} \\
&&V_2^{-}=-\frac{1}{36M}\left[(F^{\prime \prime}
-\frac{1}{2}\tilde{R} F^{\prime \prime \prime}
+2\tilde{R} {F^{\prime}}^3)s
+(4 {F^{\prime}}^2+\tilde{R} F^{\prime}F^{\prime \prime})c \right] \ .
\label{V_2^-}
\end{eqnarray}

In these equations we have introduced
the following notations:
$s\equiv sin 2F(\tilde R) $, $c\equiv cos 2F(\tilde R) $,
$\tilde R \equiv \lambda_0 R$, $F\equiv F(\tilde R)$,
$F^{\prime} \equiv dF(\tilde R)/d\tilde R$ etc.

The anti\-sym\-me\-tri\-cal spin-orbit term only contributes to the
transition potential between the $NN$ and
$NN(1440)$ channels. Denoting the nucleon
state $|0>$ and the lowest breathing state (the $N(1440)$) $|1>$ we
consider the following potential components: $V_{MN}^{\pm}\equiv
<00|V^\pm|MN>$, where $M,N = 0,1$:
\begin{eqnarray} &&V_{00}^{+} = V_0
+ 2 <0|\xi^2|0>V_2^+\ ,       \label{averV+00} \\
&&V_{10}^{+} =
V_{01}^{+} = <0|\xi^2|1>V_1^+\ ,  \label{averV+10} \\
&&V_{11}^{+} =
\left(<0|\xi|1>\right)^2 V_3\ ,   \label{averV+11} \\
&&V_{10}^{-} =
-V_{01}^{-} = <0|\xi^2|1>V_1^-\ .
\label{averV-10}
\end{eqnarray}
Here
$V_{00}^+$ is then the diagonal spin-orbit interaction in the
nucleon-nucleon channel. In the approximation
where the breathing excitations are frozen $(\lambda_1=\lambda_2=1)$
the antisymmetrical potential $V_{-}$ vanishes and the expression
(\ref{V_LS}) reduces to the result of ref.\cite{RiskaNyman}.\\

The potentials (\ref{averV+00}-\ref{averV-10})
are shown in Figs. 1--3. The result in Fig. 1 shows that the breathing
modes do not
lead to any appreciable net modification
of the isospin dependent component of
the nucleon-nucleon spin-orbit interaction at long distances
beyond 1 fm, but that at short distances their effect is
strong enough to imply a change of sign.
The short range behaviour of the results in the figures
should however be viewed as tentative
in view of the limited reliability of the product
approximation at short range,
and the omission of the effect of the stabilizing
term in the Lagrangian density in these results.
The large effect of the breathing modes
on the spin-orbit interaction at short distances
is a reflection of the softness against
radial vibrations of the skyrmions, which is most obvious in
the large underprediction of the mass of the Roper resonance
\cite{Hajduk,BDT}. \\

The results for the spin orbit components of the
transition potentials in Figs. 2
and 3 also reveal the short range of these interactions,
as well as the very large effect of the breathing modes
on the interactions at short range.\\[1.5cm]

{\bf Acknowledgement.} One of authors (A.P.K.) thanks the
Research Institute for Theoretical Physics at the
University of Helsinki for
its hospita\-li\-ty during his stay in Helsinki.\\

\vspace{2cm}

{\bf Figure Captions}

\vspace{2cm}
{\bf Figure 1.} The isospin dependent nucleon-nucleon
spin-orbit potential.
The dotted line shows breathing mode correction $V_0$ and full
line the total spin-orbit interaction $V_{00}^+$.
The potential is given in MeV; at $R=1.2\ fm$ the scale for
the potential is changed.

{\bf Figure 2.} The
spin-orbit transition potentials $V_{10}^+$ (dotted line) and
$V_{10}^-$ (full line). The potential is given in MeV; at $R=1.2\ fm$
the scale for the potential is changed.

{\bf Figure 3.} The spin-orbit transition potential $V_{11}^+$.
The potential is given in MeV.

\end{document}